# POLARIZATION ROTATION AND CASIMIR EFFECT IN SUSPENDED GRAPHENE FILMS

M. Bordag[1], I. V. Fialkovsky[2,3] (*), D. M. Gitman[3], D. V. Vassilevich[3,4]

[1] Institut fur Theoretische Physik, Universitat Leipzig, Germany
[2] Instituto de Fisica, Universidade de Sao Paulo, Brazil
[3] Department of Theoretical Physics, St. Petersburg State University, Russia
[4] Universidade Federal ABC, Sao Paulo, Brazil

(*) Corresponding author: ifialk@gmail.com

**Abstracts**

The low-energy quasi-excitations in graphene are known to be described as Dirac fermions in 2+1 dimensions. Adopting field-theoretical approach we investigate the interaction of these quasi-particles with 3+1 dimensional electromagnetic field focusing on the optical properties of suspended graphene layers and their Casimir interaction with ideal conductor. The magnitude of predicted effects (the rotation of polarization of light and the Casimir force) appears to be well within modern experimental capabilities.

## 1. Introduction

Graphene is a (quasi) two dimensional hexagonal lattice of carbon atoms. At present, it belongs to the most interesting materials in solid state physics in view of its exceptional properties and its potential applications in nano technology.

Quantum Field Theory proved to be a very successful approach to description of dynamics of elementary excitations in graphene which are supposed to be 2+1 dimensional (quasi) relativistic Dirac fermions [1,2].

In the present paper we first investigate the optical properties of graphene and show that the parity odd (quantum) effects should lead to the polarization rotation of the electromagnetic waves passing through suspended samples of mono- and few-layer graphene films. We estimate the order of the effect and reveal its quantization in magnetic field.

Secondly, we research on the Casimir interaction between a graphene sample and ideal conductor. We consider the same geometry as in the original Casimir effect (2 parallel planes) with one plane being graphene and the other one – ideal conductor.

## 2. Methodology

In the low energy approximation (i.e. in continuous limit) the quasi--particles in single layered graphene can be described as relativistic 2+1 dimensional Dirac fermions [1,2]. Its interaction with 3+1 dimensional electromagnetic field $A$ is described in gauge invariant way

$$S_\psi = \int d^3x\, \bar\psi (i\tilde{\not\partial} - e\tilde{\not A} + \ldots)\psi, \quad S_{EM} = \int d^4x\, F_{\mu\nu}^2$$

$$\tilde{\not\partial} = \tilde\gamma^l \partial_l,\ l=0,1,2,\ \mu,\nu = 0,1,2,3, \quad (1)$$

$$\tilde\gamma^0 = \gamma^0,\ \tilde\gamma^{1,2} = v_F \gamma^{1,2},\ (\gamma^0)^2 = -(\gamma^{1,2})^2 = 1$$

The Dirac matrices γ may be taken in a reducible representation, $v_F$ is the Fermi velocity. We assume $\hbar = c = 1$, $v_F \approx (300)^{-1}$, $e^2 = 4\pi\alpha = 4\pi/137$.

Any other parameters that can enter the model, like the mass or chemical potential, are denoted with the dots in the action, see [3] for a list of possible interactions.

The system as a whole is described by generating functional

$$Z = N\int DA D\psi D\bar\psi\, e^{i(S_\psi + S_{EM})} \approx N'\int DA\, e^{i(S_{EM} + S_{eff})}$$

where to obtain the last expression we integrated out the fermions and retained only the quadratic order in $A$ of the effective action for the electromagnetic potential

$$S_{eff} = A \,\triangleleft\!\bigcirc\!\triangleright\, A = \frac{1}{2}\int d^3x\, d^3y\, A_j(x)\Pi^{il}(y-x)A_l(y).$$

The polarization operator is given by

$$\Pi^{mn}(x) = \int \frac{d^3p}{(2\pi)^3} e^{ipx}\Pi^{mn}(p), \quad (2)$$

$$\Pi^{mn}(p) = \frac{\alpha}{v_F^2}\eta_j^m\left[\Phi(\tilde{p})\left(g^{jl} - \frac{\tilde{p}^j\tilde{p}^l}{\tilde{p}^2}\right) + i\phi(\tilde{p})\epsilon^{jkl}\tilde{p}_k\right]\eta_l^n$$

with $\varepsilon^{012} = 1$, $\eta_j^n = \text{diag}(1, v_F, v_F)$, $\tilde{p}^m = \eta_j^m p^j$.

The functions $\Phi$ and $\phi$ are model-dependent and can take complex values, the parity-odd part can originate either due to the chiral anomaly inherent to 2+1 dimensional fermions [1] or due to the presence of external (magnetic) field [3]. Due to the presence of effective 2-dimensional speed of light $v_F$, $\Pi_{\mu\nu}$ depends on the rescaled momentum $\tilde{p}$. The multiplier $v_F^{-2}$ appears due to the relation $d^3p = v_F^{-2}d^3\tilde{p}$ for the integration measure for the loop momentum. The overall rescalings $\eta_j^n$ of the polarization operator appear since the electromagnetic potential is also multiplied with rescaled gamma-matrices $\tilde{\gamma}$. The polarization tensor is transversal (gauge invariant) with respect to the unrescaled momenta, $p_m\Pi^{mn}(p) = 0$. For explicit calculations of functions $\Phi, \phi$ and related issues see [1,3].

Basing on this field theoretical approach we calculate two potentially observable effects.

## 3. Results & Discussion

### 3.1 Faraday rotation

The modified Maxwell equations corresponding to the effective action are

$$\partial_\mu F^{\mu\nu} + \delta(z)\Pi^{\nu\rho}A_\rho = 0$$

which lead to the following matching condition imposed on the field $A_\mu$

$$A_\mu\big|_{z=+0} = A_\mu\big|_{z=-0},$$
$$(\partial_z A_\mu)_{z=+0} - (\partial_z A_\mu)_{z=-0} = \Pi^{\nu\rho}A_\rho\big|_{z=0}$$

For linearly $x$-polarized plane waves propagating along the $z$-axis from $z = -\infty$, the amplitude of the transmission $\mathbf{A}$ and angle of polarization rotation $\theta$ read [4]

$$\mathbf{A} \simeq 1 - \left|\frac{\text{Im}\,\Phi}{2\omega}\right|\alpha + O(\alpha^2), \quad \theta \simeq -\frac{\alpha\,\text{Re}\,\phi}{2} + O(\alpha^2)$$

For the simplest model defined by (1) the functions $\Phi$ and $\phi$ are given by

$$\Phi = N\frac{2m\tilde{p} - (\tilde{p}^2 + 4m^2)\text{arctanh}(\tilde{p}/2m)}{2\tilde{p}}, \quad (4)$$
$$\phi = \frac{2m\,\text{arctanh}(\tilde{p}/2m)}{\tilde{p}} - 1$$

Then the absorption of visible light is in agreement with experiment [5] at the linear order in $\alpha$.

Polarization rotation in magnetic field $\mathbf{B}$, i.e. Faraday effect, can also be investigated in non-zero temperature $T$ and non-vanishing chemical potential $\mu$. Basing on the explicit calculations of polarization tensor (or conductivity) [3] one obtains

$$\phi(\omega) = \frac{\varepsilon_B L_b^2(\omega^2 - L_b^2(1+n_0))}{\omega^4 + L_b^4 - 2\omega^2 L_b^2(1+n_0)}, \quad n_0 = \left[\frac{\mu^2}{L_b^2}\right]$$

where $L_b^2 = 2v_F^2|e\mathbf{B}|$. This leads to the quantization of Faraday rotation of polarization of light passing through suspended graphene films. The results of such quantization are exemplified at Fig. 1.

Faraday effect and its quantization in graphene was envisaged in [6], theoretically predicted in [4] and numerically investigated in [7].

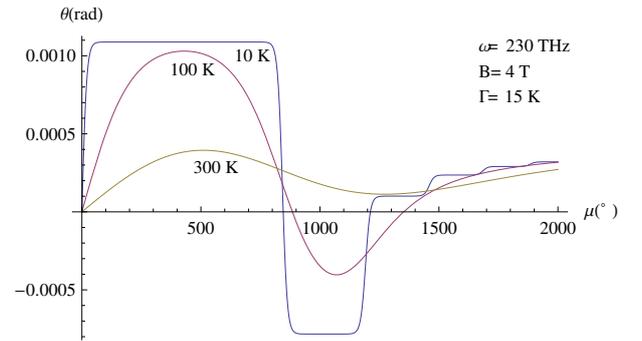

Fig. 1. Polarization rotation angle $\theta$ depending on the chemical potential $\mu$.

### 3.2 Casimir energy

Let us consider an interaction of a graphene plane (at $x^3 = 0$) and parallel ideal conductor (at $x^3 = a$). At the leading order in the fine structure constant $\alpha$ the Casimir energy is given by

$$E = \frac{-i}{TS}\text{Log}\, Z \approx -\frac{1}{TS} \bigcirc + O(\alpha^2) \quad (5)$$

The wave line here denote the photon propagator $D_{\mu\nu}$, and in the Feynman gauge is diagonal. Since $\Pi^{ij}$ by construction does not have components along the $x^3$ direction, we are only interested in the $D_{ij}$ part which satisfy Dirichlet boundary conditions, and one can write

$$D_{ij}(x,y) = g_{ij}\left(D_0(x-y) - D_0(x-y_R)\right)$$

$$D_0(x) = \int \frac{d^3 p}{(2\pi)^3} e^{ip_j x^j} \frac{e^{-p_\parallel |x^3|}}{2 p_\parallel}, \quad p_\parallel \equiv |p|$$

In calculation of the bubble diagram in (5) the term with $D_0(x-y)$ will be neglected as it does not contribute to the Casimir force being independent on $a$. Thus after the Fourier transformation and the Wick rotation we obtain for the Casimir energy [8,9]

$$E = -\frac{1}{4}\int \frac{d^3 p_E}{(2\pi)^3} \frac{\Pi^j_j(p_E)}{p_\parallel} e^{-2ap_\parallel} = \\ -\frac{\alpha}{4}\int \frac{d^3 p_E}{(2\pi)^3} \frac{(p_\parallel^2 + \tilde{p}_\parallel^2)\Phi(p_E)}{p_\parallel \tilde{p}_\parallel^2} e^{-2ap_\parallel} \quad (6)$$

where we expanded $\Pi^{ij}(p_E)$ explicitly with help of (2).

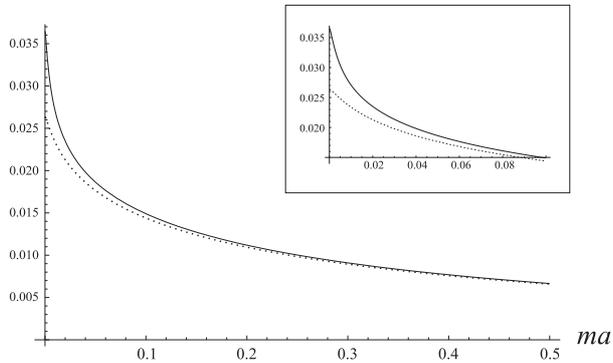

Fig. 2. Casimir energy normalized to ideal case $E_C = -\frac{\pi^2}{720 a^3}$.

Analytical asymptotic of the energy could also be obtained and are given by

$$E \underset{a\to\infty}{\approx} -\frac{\alpha N}{96\pi^2}\frac{2+v_F^2}{ma^4}, \quad E \underset{a\to 0}{\approx} -\frac{\alpha}{16\pi a^3} h(N, v_F).$$

## 4. Conclusions

We showed that the presence of parity-odd terms in the polarization tensor of Dirac quasi-particles (i.e. in the conductivity tensor) leads to rotation of polarization of the electromagnetic waves passing through suspended graphene films. In external magnetic field this gives rise to a quantum Faraday effect. In this case the rotation angle gets quantized in low-temperature limit. The estimated order of the effect is well above the sensitivity limits of modern optical instruments.

We also presented the calculation of the Casimir interaction energy between a suspended graphene sample and a parallel plane perfect conductor. The Casimir interaction for this system appears to be rather weak though potentially measurable. It exhibits strong dependence on the mass of the quasi-particles in graphene.